\documentclass{article}
\usepackage{amssymb}
\usepackage{amsmath}
\usepackage{hyperref}
\usepackage{color}

\input{tcilatex}
\begin{document}

\begin{center}
\textbf{Random-coefficient pure states, the density operator formalism and
the Zeh problem}

Alain Deville

\dag Aix-Marseille Universit\'{e}, CNRS, IM2NP UMR 7334, F-13397 Marseille,
France

Yannick Deville

*Universit\'{e} de Toulouse, UPS, CNRS, CNES, OMP, IRAP (Institut de
Recherche en Astrophysique et Plan\'{e}tologie),

F-31400 Toulouse, France

\ \ \ 
\end{center}

\textbf{Abstract }

Quantum electronics is significantly involved in the development of the
field of quantum information processing.\ In this domain, the growth of
Blind Quantum Source Separation and Blind Quantum Process Tomography has
led, within the formalism of the Hilbert space, to the introduction of the
concept of a Random-Coefficient Pure State, or RCPS: the coefficients of its
development in the chosen basis are random variables. This paper first
describes an experimental situation necessitating its introduction.\ While
the von Neumann approach to a statistical mixture considers statistical
properties of an obser\-vable, in the presence of an RCPS one has to
manipulate statistical properties of probabilities of measurement outcomes,
these probabilities then being themselves random variables. It is recalled
that, in the presence of a von\ Neumann statistical mixture, the consistency
of the density operator $\rho $ formalism \ is based on a postulate. The
interest of the RCPS concept is\ presented in the simple case of a spin $1/2$%
, through two instances. The most \ frequent use of the $\rho $ formalism by
users of quantum mechanics is a motivation for establishing some links
between a given RCPS and the\ language of the density operator formalism,
while keeping in mind that the situation described by an RCPS is different
from the one which has led to the introduction of $\rho $. It is established
that the Landau - Feynman use of $\rho $ is mobilized in a situation
differing from both the von Neumann statistical mixture and the RCPS. It is
shown that the use of the higher-order moments of a well-chosen random
variable helps solving a problem already identified by Zeh in 1970.

\section{Introduction\label{SectionIntroduction}}

Superconducting qubits are presently proposed as a possible solution in the
building of quantum gates. John Bardeen was twice a Nobel Laureate, in 1956
with Shockley and Brattain for their invention of the transistor, and in
1972 with \ Cooper and Schrieffer for their theory of superconductivity. In
both cases Quantum Mechanics (QM) was mobilized. The developments of
Telecommunications and Electronics have led to the birth and growing of a
Theory of Information, first in the classical context (see e.g. the
appearance of the Shannon entropy \cite{Shannon1948}) and, for several
decades, in the quantum domain (see e.g. the \textit{Feynman\ Lectures on
Computation} \cite{Feynmann1996OnComputation},\textit{\ }and \textit{Quantum
Computation and Quantum Information }by Nielsen and Chuang \cite%
{NielsenChuang2005}). Quantum Information Processing (QIP) is a significant
part of the Quantum Information field, and the development of quantum gates
and more generally of quantum circuits devoted to QIP is an important
activity within Quantum Electronics, itself reflected in the existence of
the Quantum Electronics Section in this Journal.

Working in the field of QIP for more than fifteen years, we have been led
first to extend the classical field of Blind Source Separation (BSS) \cite%
{Comon2010},\cite{DevilleYJutten2010} to a quantum version, namely Blind
Quantum Source Separation (BQSS) \cite{DevilleYDevilleAICA2007}, \cite%
{DevilleYAQIP2012}, \cite{DevilleYANaik2014}, \cite{Entropy2017}. More
recently we introduced the field of Blind Quantum Process Tomography (BQPT) 
\cite{DevilleYABQPTLiberec2015}, \cite{DevilleYAToulouse2017}, \cite%
{DevilleYAPRA2020}, \cite{Entropy2017}, an extension of Quantum Process
Tomography (QPT). In these contexts, we were led not to use the density
operator formalism, but to introduce what, as in \cite{DevilleY2023QIP}, is
hereafter called a Random-Coefficient Pure State (RCPS): $\Sigma $ being an
isolated quantum system, $\mathcal{E}$ its state space, with dimension $d,$
and \{$\mid k>$\} an orthonormal basis of $\mathcal{E},$ it is considered
that at\textit{\ some time }$t_{r}$ $\Sigma $ may be in a random-coefficient
pure state 
\begin{equation}
\mid \Psi >=\sum_{k}c_{k}\mid k>,\text{ \label{Eq. RCPS developpe}}
\end{equation}%
where the $c_{k}$ are Random Variables (RV), with the constraint\textbf{\ }$%
\sum_{k}\mid c_{k}\mid ^{2}=1$\textbf{.}\ \ In contrast, the coefficients of
the development of the usual pure states are deterministic quantities. It
must be emphasized from the beginning that if \{$\mid k>$\} is an eigenbasis
of some observable $O$ attached to $\Sigma $, then if the state of $\Sigma $
is described by this RCPS $\mid \Psi >$ and if $O$ is measured,\textit{\ the
probability} of obtaining the (assumed non-degenerate) eigenvalue associated
with $\mid k>$, i.e. $\mid c_{k}\mid ^{2},$ \textit{is itself an RV.}

The present paper uses standard Quantum Mechanics (QM). As a result of its
postulates, including the existence of a principle of superposition (of
states), which the late Nobel Laureate Steven Weinberg called the first
postulate of QM \cite{Weinberg2013}, then, given a quantum system $\Sigma $,
and its state space $\mathcal{E}$, a Hilbert space, any vector of $\mathcal{E%
}$ (defined up to a phase factor $e^{i\varphi }$, $\varphi $ being a real
quantity) represents a possible state of $\sum $ called a pure state. This
standard Hilbert space framework is used by both the so-called orthodox
interpretation of QM (Bohr, Heisenberg, Pauli, Rosenfeld) and by the
statistical interpretation (Einstein, Schr\"{o}dinger, Blokhintshev,
Ballentine), with the meaning given by Ballentine \cite{Ballentine1970} to
that latter expression, one of these interpretations more or less implicitly
accepted by many users of QM. Weinberg has stressed that "\textit{quantum
field theory is based on the same quantum mechanics that was invented by Schr%
\"{o}dinger, Heisenberg, Pauli, Born, and others in 1925-26, and has been
used ever since in atomic, molecular, nuclear, and condensed matter physics" 
}\ \cite{WeinbergQuantumFieldsV1}\textit{. }We warn the reader that we are
therefore outside the approach initiated by Segal \cite{Segal1947}, with his
introduction claiming "\textit{Hilbert space plays no role in our theory}",
an approach known as the C*-algebra formulation of QM, then developed by
Haag and Daniel Kastler, and more recently by Strocchi \cite{Strocchi2012}
(see also \cite{Drago2018}). We are also outside the approach from Mielnik 
\cite{Mielnik} and again Haag, together with Bannier \cite{HaagNonLinearQM}.
The formal constructions and possible results from mathematical physicists
trying to build general quantum theories aiming at unifying general
relativity and Quantum Mechanics (QM), an important field in present day
Physics, are out of the scope of this paper.

A state of the Hilbert space - pure state - used in QM, and described by a
ket in the Dirac formalism, obeying the Schr\"{o}dinger equation if $\Sigma $
is isolated,\ can be obtained from a preparation act. von Neumann considered
a more general situation, called a mixed state or statistical mixture (of
states), and, as a consequence of\ his introduction of a postulate (cf.\
Sections \ref{SectionvonNeumannMixture-Postulate} and \ref%
{SectionTheZehProblem}),\ established that it can be formally described with
a density operator $\rho $ \cite{vonNeumann1927}, \cite{vonNeumann1932}.

An experimental situation leading to the introduction of an RCPS is first
described in Section \ref{SectionExistenceRCPS}, and the ambiguity in the
present use of the expression random pure state is stressed.\ Within a given
theory (an abbreviation for \textit{a given theoretical frame, }\ here the
standard version of quantum mechanics), one should obviously distinguish
between an experimental (or possibly simulated) situation and the formal
tools used in its description. One should therefore keep in mind the origin
of a given formal tool when deciding to use\ it in a given situation. 
\textit{An RCPS is not a statistical mixture, and our already cited papers
devoted to either BQSS\ or BQPT, including our recent paper \cite%
{DevilleY2023QIP}, did not use the density operator }$\rho $\textit{.
However, since }$\rho $\textit{\ is quite often used in QM studies, it is
useful to establish links between the experimental situation described by an
RCPS and the formal density operator, }clearly distinguishing between what
is related to an RCPS and the possible suggestions about the content of the
density operator $\rho $ in the specific situations described by a
statistical mixture. The present work is \textit{not} devoted to a
historical study, but can't ignore that the present debates within QM are
still largely dependent upon its developments in the 1924-1935 years.

In Section \ref{SectionvonNeumannMixture-Postulate}, the existence of the
measurement postulate introduced by von Neumann and its consequences in the
description of statistical mixtures are presented. Links between an RCPS and
the density operator $\rho $ are established in Section \ref%
{SectionLinksRCPSAndRho}. The interest of the RCPS concept is\ presented in
Section \ref{SectionInterestRCPSTwoInstances} in the simple case of a spin $%
1/2.$ In a first instance, the RCPS depends on a single real random
parameter obeying a truncated Gaussian law, with two unknown real
parameters, and one accesses experimental or simulated data. It is
impossible to evaluate these two unknown parameters through the density
matrix associated with this RCPS, but this can be done, within this quantum
context, thanks to statistical properties of a probability, not of an
observable.\ A second instance numerically compares two RCPS with the same
associated density operator, and which are shown to be different when using
also a moment with higher order than the one used through the density
operator formalism. The reader may consult our already cited papers for
quite more complex uses of the RCPS concept. In Section \ref%
{SectionLandau-Feynman}, it is explained that the limitations of the $\rho $
formalism also exist in a situation historically first discussed by Landau
and later on clarified by Feynman, and which, for brevity, we call the
Landau-Feynman approach. In Section\ \ref{SectionTheZehProblem}, stimulated
by our use of higher-order moments in the presence of an RCPS, we come to
the Zeh problem, with a spin 1/2 and \textit{the two von Neumann mixed
states considered by Zeh}, described by the same density operator. Using the
moments of an RV linked to the results of measurements, we show that if the
measured spin component and the RV are both well-chosen, the values of 
\textit{at least one of its moments} differ, when considering these two
mixtures, which allows us to differentiate between these statistical
mixtures. Conclusions are drawn in Section \ref{SectionConclusion}.

\section{An experimental situation with a system in an RCPS\label%
{SectionExistenceRCPS}}

We\ recall a simple situation - detailed, in the context of BQSS, in \cite%
{Entropy2017} - when the random-coefficient pure state concept is
meaningful. $\Sigma $ consists of the magnetic moment $\overrightarrow{\mu }$%
\ of an electron spin $1/2$, with $\overrightarrow{\mu }$\ $=-G$\ $%
\overrightarrow{s}$\ (isotropic $\overline{\overline{g}}$\ tensor), in a
static field $\overrightarrow{B_{0}}$=$B_{0}\overrightarrow{Z}$\ with
amplitude $B_{0}.$\ Writing the Zeeman Hamiltonian as $h=-\overrightarrow{%
\mu }\overrightarrow{B}_{0}$\ = $GB_{0}s_{Z}$\ indicates that while the spin
is a quantum object, the magnetic field is treated classically. Someone (the
Writer) first prepares the spin in the 
\TEXTsymbol{\vert}$
+Z\rangle $\ eigenstate of $s_{Z}$\ (eigenvalue $+1/2$). The moment is then
received by a second person (the Reader), who ignores the direction of $%
\overrightarrow{B_{0}}$, chooses some direction $z$\ (unit vector $%
\overrightarrow{u_{z}}$) \ attached to his own Laboratory as the
quantization direction and introduces a Laboratory-tied cartesian frame $xyz$%
, used to define $\theta _{E}$\ and $\varphi _{E}$, the Euler angles of $%
\overrightarrow{Z}$. Since the field is treated classically, $\theta _{E}$\
and $\varphi _{E}$\ behave as classical variables, while $s_{Z}$\ is an
operator. The Reader measures $s_{z}$\ $=\overrightarrow{s}\overrightarrow{%
u_{z}}$\ (eigenstates: 
\TEXTsymbol{\vert}$
+\rangle $\ and 
\TEXTsymbol{\vert}$
-\rangle $), and is interested in the probability $p_{+z}$\ of getting $+1/2$%
. An elementary calculation indicates that, when the time interval between
writing and reading may be neglected:%
\begin{equation}
\text{\TEXTsymbol{\vert}}+Z\rangle =\alpha \text{\TEXTsymbol{\vert}}+\rangle
+\sqrt{1-\alpha ^{2}}e^{i\varphi }\text{\TEXTsymbol{\vert}}-\rangle ,\text{ %
\label{+ZAvecRetPhi}}
\end{equation}%
with\textbf{\ }%
\begin{equation}
\alpha =\cos \frac{\theta _{E}}{2},\qquad \varphi =\varphi _{E},
\end{equation}%
and therefore $p_{+z}=\cos ^{2}\theta _{E}/2.$ Once the direction of $%
\overrightarrow{B_{0}}$\ has been chosen, state 
\TEXTsymbol{\vert}$
+Z\rangle $\ is then unambiguously defined. If this direction has a
deterministic nature, $%
\alpha 
$\ and $\varphi $\ are deterministic variables, and 
\TEXTsymbol{\vert}$
+Z\rangle $,~usually called a pure state, may be called a
deterministic-coefficient pure state. If $\theta _{E}$\ and $\varphi _{E}$\
obey probabilistic laws, one may consider that the quantum quantities $%
\alpha 
$\ and $\varphi $, which depend upon the classical RV $\theta _{E}$\ and $%
\varphi _{E}$, do possess the properties of conventional, i.e. classical,
RV. We are not strictly facing the quantum equivalent of a classical
situation here. Rather, the stochastic character of the field direction,
with classical nature, is reflected in the random behaviour of the quantum
state expressed through Eq. (\ref{+ZAvecRetPhi}). While random operators are
well known e.g. in NMR (see Ch. VIII of \cite{Abragam1961RMN}), we here meet
a random-coefficient pure state. And the probability $p_{+z},$ equal to $%
\cos ^{2}\theta _{E}/2\,,$ is therefore itself a random variable.

In the field of probability theory, a vector whose components are random
variables is called a random vector (see e.g. page 243 of \cite%
{PapoulisPillai}). $\ $In Eq. (\ref{+ZAvecRetPhi}), $\mid +Z\rangle ,$ with
its random coefficients $%
\alpha 
$ and $\varphi ,$ may therefore be called a random ket, or as describing a
random pure state (of the spin $1/2$), and since our 2007 paper \cite%
{DevilleYDevilleAICA2007}, we used these expressions with this meaning,
keeping in mind that, once a given sample of this random vector has been
selected, then its components, in the considered quantum context, while
being deterministic quantities, have a probabilistic content. In order to
try and suppress any ambiguity, in the present paper, as in \cite%
{DevilleY2023QIP}, instead of speaking of a \textit{random pure state} we
speak of a \textit{random-coefficient pure state}, since moreover the
expression \textit{random quantum pure states} was already used in 1990 by
Wootters \cite{Wootters1990} with three different contents, and since, as
detailed in Section 5 of \cite{DevilleY2023QIP}, the expression \textit{%
random pure states }today appears with different meanings.\ 

\section{von Neumann statistical mixture and measurement postulate\label%
{SectionvonNeumannMixture-Postulate}}

With a physical observable quantity $O$ attached to $\Sigma $, QM associates
a linear Hermitian operator $\widehat{O}$ acting on the states of $\Sigma .$%
\ The mean value of $\widehat{O}$ when $\Sigma $ is in the pure (normed)
state $\mid \Psi >$ is a quantity\ written, in the Dirac formalism, as $%
<\Psi \mid $ $\widehat{O}\mid $ $\Psi >$. A statistical mixture, as
historically introduced by von Neumann \cite{vonNeumann1927}, \cite%
{vonNeumann1932}, is denoted as \{$p_{i}$, $\mid \varphi _{i}>$\}, where $%
p_{i}$ is the probability of presence of the normed pure state$\ \mid
\varphi _{i}>$ \ (for any $i,$\ $p_{i}\geq 0$ and $\sum_{i}p_{i}=1$), and
the mean value of $\widehat{O}$ is then equal to $\sum_{i}p_{i}<\varphi
_{i}\mid \widehat{O}\mid \varphi _{i}>$. With such a statistical mixture,
one associates its density operator $\rho =\Sigma _{i}p_{i}\mid \varphi
_{i}><\varphi _{i}\mid ,$ acting linearly on the (hereafter assumed to be
normed) states of $\Sigma $. The eigenvalue spectrum of a Hermitian positive
operator with finite trace is entirely discrete, a result of Hilbert space
theory (\cite{Messiah1}, page 335). When an isolated system is in a
statistical mixture, $\rho $ obeys the Liouville-von Neumann equation. \ In
the special case when $\Sigma $ is in a pure state $\mid \Psi >,$ $\rho $ is
a projector: $\rho =\mid \Psi ><\Psi \mid .$ The relation $Tr\rho ^{2}\leq
Tr\rho $ is obeyed by $\rho $, the equality being verified \ iff $\rho $ is
a projector, i.e. if and only if $\rho $ describes a pure state. A \textit{%
fundamental postulate} (see e.g. Peres \cite{Peres1995}, pages 75-76, and
Section \ref{SectionTheZehProblem}), first proposed by von Neumann, is used:
"\textit{the }$\rho $\textit{\ matrix completely specifies all the
properties of a quantum ensemble" }\cite{Peres1995}. Then, $\widehat{O}$
being attached to an arbitrary observable $O$ of $\Sigma $, von Neumann
claims that the whole information which can be reached through measurements
of $O$ is contained in the expression $E\{\widehat{O}\}=Tr\{\rho \widehat{O}%
\}$ (von\ Neumann introduced the density operator $\rho $ while aiming at
this result, cf.\ Section \ref{SectionTheZehProblem}).\ Consequently, the
writing \{$p_{i}$, $\mid \varphi _{i}>$\} is considered ambiguous: if two
so-defined statistical mixtures possess the same density operator they must
be seen as the same statistical mixture.

In the situation considered by von Neumann, the limits of the $\rho $ tool
are therefore those of a postulate, but they are not always identified, a
result of von\ Neumann's authority, as in \cite{vonNeumann1932} he wrote
that he had demonstrated a result about $\rho $ (which he named as $U$),%
\textit{\ }whereas, as shown in Section \ref{SectionTheZehProblem}, \textit{%
in fact he had just postulated it.}

\section{RCPS and the density operator $\protect\rho $ \label%
{SectionLinksRCPSAndRho}}

Starting from a given RCPS, it is shown that one and only one density
operator can be associated with it.\ Then, starting from a given density
operator,\ it is shown that more than one RCPS can be associated with it.

\subsection{From an RCPS to $\protect\rho $\label{SubsectionFromrRCPStoRho}}

\textit{At some time }$t_{r},$ an isolated system $\Sigma $ (state space $%
\mathcal{E}$, with dimension $d,$ and \{$\mid k>$\} an orthonormal basis of $%
\mathcal{E}$) is supposed to be in an arbitrary random-coefficient pure state%
\begin{equation}
\mid \Psi >=\sum_{k}c_{k}\mid k>,\text{ \label{PsiRCPSdeveloppe}}
\end{equation}%
where the $c_{k}$ are RV, with the constraint\textbf{\ }$\sum_{k}\mid
c_{k}\mid ^{2}=1$\textbf{.}\ One is interested in the mean value then taken
by the scalar Hermitian operator $\widehat{O}$ associated with some
observable of $\Sigma $. One first considers a given choice of the value of
each RV\ $c_{k}.$\ The contribution of this specific, then
deterministic-coefficient pure state, denoted as $\mid \Psi _{s}>$, is, from
the rules of QM:%
\begin{equation}
<\Psi _{s}\mid \widehat{O}\mid \Psi _{s}>=\sum_{k,l}c_{k}^{\ast }<k\mid 
\widehat{O}\mid l>c_{l}.
\end{equation}%
The mean value of $\widehat{O}$ when $\Sigma $ is in this RCPS is defined as
the expectation (denoted as $E$) or mean value of this quantity:%
\begin{equation}
E\{<\Psi _{s}\mid \widehat{O}\mid \Psi _{s}>\}=\sum_{k,l}E\{c_{k}^{\ast
}c_{l}\}\widehat{O}_{kl}=\sum_{k,l}r_{lk}\widehat{O}_{kl}=Tr(r\widehat{O}),
\end{equation}%
where, in the chosen basis, $\widehat{O}_{kl}=<k\mid \widehat{O}\mid l>$ and%
\begin{eqnarray}
r_{lk} &=&<l\mid r\mid k>=E\{c_{k}^{\ast }c_{l}\}\text{ \label{r indice lk}}
\\
r &=&\sum_{k,l}r_{kl}\mid k><l\mid .\text{\label{matrice r}}
\end{eqnarray}%
The introduction of the linear operator $r$ with matrix elements $r_{lk}$\
in the \{$\mid k>$\} basis, and that of a trace, are the consequences of the
superposition principle and of the fact that the expectation of the sum is
equal to the sum of the expectations. $r$ has the following properties,
resulting from a transposition of the usual arguments in the presence of a
statistical mixture to the present situation, i.e. an arbitrary RCPS (cf.
also the Appendix):

1) for any pair $(k,l)$, $r_{lk}$ $=r_{kl}^{\ast }$:\ the operator $r$ is
Hermitian,

2) in a basis \{$\mid i>$\} in which the Hermitian operator $r$ is diagonal,
its diagonal elements are its eigenvalues; being of the form $E\{\mid
c_{i}\mid ^{2}\}=p_{i},$\ they are all non-negative ($r$ is a positive
operator), and their sum is equal to $1$,

3) for any ket $\mid u>,$\ $<u\mid r\mid u>\geq 0,$\ since it is equal to $%
\sum_{i}p_{i}\mid <i\mid u>\mid ^{2}$,

4) $H$\ being the Hamiltonian of $\Sigma ,$\ for each choice of the $c_{k}$\
values the corresponding ket $\mid \Psi _{s}>$\ obeys the Schr\"{o}dinger
equation, and from $t_{r}$ on,$\ r$\ obeys the Liouville-von Neumann
equation $i\hslash dr/dt=[H,r].$

\ This operator $r$ has all the properties of a density operator. This
result had been suggested, for a qubit, in \cite{Entropy2017}.

It must be realized that the Gleason theorem can be opposed neither to the
existence of an RCPS nor to the establishment of the properties discussed in
this section, since the establishment of this theorem starts by introducing
projectors (see e.g \cite{LeBellac2007}, page 189), i.e. a formal object
associated with deterministic-coefficient pure states.

\subsection{From a given $\protect\rho ,$ towards RCPS\label%
{SubsectionFromRhoToRCPS}}

It has been recalled that, as a consequence of the von Neumann postulate
(cf. Section \ref{SectionvonNeumannMixture-Postulate}), if several
statistical mixtures \{$p_{i}$, $\mid \varphi _{i}>$\} have the same density
operator, they are considered as being the same statistical mixture. But a
statistical mixture is not an RCPS. Presently starting with system $\Sigma $
and its state space $\mathcal{E},\Sigma $ is assumed to be in a statistical
mixture described by a density operator $\rho $.\ In an orthonormal basis of 
$\mathcal{E},$ \{$\mid i>$\}, in which $\rho $ is diagonal:%
\begin{equation}
\rho =\sum_{i}p_{i}\mid i><i\mid
\end{equation}%
where the $p_{i}$ are probabilities. Is it possible to associate at least
one and possibly more than one RCPS with $\rho ?$ It is hereafter shown that
the answer is yes.

We are first able to show that at least some well-chosen RCPS $\mid \Psi
_{1}>=\sum_{i}c_{i}$ $\mid i>,$ developed over this basis and the $c_{i}$
being RV, may be associated with a statistical matrix $r$ equal to that
representing $\rho $ in that basis, even while the following strong
conditions have been imposed upon the coefficients $c_{i}$: 1) the RV\ $%
c_{i} $ are real, which avoids considering complex RV.\ 2) RV $c_{d}$ obeys $%
c_{d}=\delta \sqrt{1-\sum_{i=1}^{d-1}c_{i}^{2}},$ $\delta $ being an RV
taking the values $+1$ and $-1,$ each with probability $1/2$, and $\delta $
is statistically independent from all the $c_{i}$ with $i<d.$ 3)$\ $All the $%
c_{i},$ $c_{j\neq i}$ pairs with $i<d$ and $j<d$ are statistically
independent. 4) each RV $c_{i}$ with $i<d$ obeys a centered, truncated ($%
\mid c_{i}\mid \leq 1\mid $) Gaussian probability density function, with a
variance equal to $p_{i}.$ As a consequence of these assumed properties, if $%
i<d,$ $j<d$ and $i\neq j,$ then $r_{ij}=$ $<c_{j}><c_{i}>=0,$ and if $i<d$
and $j=d$, then $r_{id}=<c_{i}\delta \sqrt{1-\sum_{l=1}^{d-1}c_{l}^{2}}>$
and, as\ $\delta $ is independent from the $c_{i}$ and is a centered RV, $%
r_{id}=0;$ the same is true for $r_{dj}$ with $j<d.$\ Therefore, the
non-diagonal matrix elements of $r,~$the statistical matrix associated with
this random-coefficient pure state in the chosen basis, namely the mean
values $r_{ij}=<c_{j}^{\ast }c_{i}>$, with $j\neq i,$ are all equal to $0$:
in the orthonormal basis \{$\mid i>$\} both $r$ and $\rho $ are diagonal
matrices. And, from 4), $r_{ii}=p_{i}.$\ Therefore, starting from a
statistical mixture described by a density operator $\rho ,$ it has been
possible to\ build a random-coefficient pure state with a matrix $r$ equal
to $\rho .$ We will say that $\mid \Psi _{1}>$ may be associated with $\rho
. $

A second possible random-coefficient pure state $\mid \Psi _{2}>,$ again
written as in Eq. (\ref{PsiRCPSdeveloppe}) (the $c_{i}$ are again RV), may
be supposed to obey assumptions 1), 2), 3), whereas 4) is now replaced by
the following condition 4'): each RV $c_{i}$ with $i<d$ has\ a centered
truncated ($\mid c_{i}\mid \leq 1\mid $) Laplace probability density
function, again with a variance equal to $p_{i}.~$Then $\mid \Psi _{2}>$ may
be associated with $\rho .$

A third possible random-coefficient pure state $\mid \Psi _{3}>$, again
written as in Eq. (\ref{PsiRCPSdeveloppe}) (the $c_{i}$ are again RV) may be
built, which is supposed to obey the following conditions: assumptions 1)
and 2) are kept, but it is now assumed that: $3^{\prime })$ if $i<d,$ $j<d$
and $i\neq j$, any mean value $r_{ij}=<c_{j}^{\ast }c_{i}>$ is equal to zero
and $4")$\ for any value of $i,$ $c_{i}$ obeys $<c_{i}^{2}>=p_{i}.$\ Then $%
\mid \Psi _{3}>$ may be associated with $\rho .$

\section{Two instances of the interest of the RCPS concept\label%
{SectionInterestRCPSTwoInstances}}

An experimental situation with an RCPS was described in Section \ref%
{SectionExistenceRCPS}. In simulations, an RCPS can be imagined at the input
of some device with only partly known properties, and from the state
obtained at its output, information can be obtained about the device itself.
We present two simple instances, using a spin 1/2, showing the interest of
the RCPS concept.\ A recent, more complex instance of the interest of RCPS,
in the domain of Blind Quantum Process Tomography, may be found in \cite%
{DevilleY2023QIP}.

\subsection{A spin 1/2 in an RCPS\label{SubsectionInstanceRCPSTwoEquations}}

One decides to consider an isotropic magnetic moment associated with a spin
1/2, in a situation when its state is described by the following RCPS:%
\begin{equation}
\mid \Phi >=\alpha \mid +>+\sqrt{1-\alpha ^{2}}\mid ->\text{ \ \label%
{spin1/2RCPS}}
\end{equation}%
with $s_{z}\mid \pm >=(\pm 1/2)\mid \pm >.$\ $
\alpha 
$ is a real RV, following a truncated Gaussian law, with $0\leq 
\alpha 
\leq 1,$ with two unknown parameters, its mean value $\eta $ and its
variance $\sigma ^{2}.$

At a first level, we introduce $N$ independent identical systems, each one
with this isotropic magnetic moment and a Stern-Gerlach device, the static
field having the same well-defined direction (denoted as $Z$) and amplitude
in all devices, and all the spins being in the same
deterministic-coefficient pure state at the input of the Stern-Gerlach
device. The $s_{z}$ component of each spin is then measured, which allows us
to get the mean value of $\ s_{z}$ for this given direction of the magnetic
field. One could also imagine a single spin and a single device, the same
experiment being made $M$ times, with $M>>1$ (more time consuming, but with
a single equipment).

At a second level, as suggested by the content of Section \ref%
{SectionExistenceRCPS}, the amplitude of the magnetic field being unchanged,
this experiment is made for (a high number of) random directions of the
magnetic field, with however systematically $\varphi =0$ in Eq. (\ref%
{+ZAvecRetPhi}), and $\alpha $ obeying the just defined truncated Gaussian
law.

One first uses the $\rho $ formalism, introducing the density operator $r$
associated with this RCPS, given by Eq. (\ref{matrice r}). The mean value of 
$s_{z}$ is then%
\begin{eqnarray}
E\{s_{z}\} &=&Tr\{rs_{z}\}=r_{++}<+\mid s_{z}\mid +>+r_{--}<-\mid s_{z}\mid
-> \\
&=&\frac{1}{2}r_{++}-\frac{1}{2}r_{--}=r_{++}-\frac{1}{2}.\text{\label%
{spin1/2DensityOperator}}
\end{eqnarray}%
$E\{s_{z}\}$ is estimated from the experiments (or the simulations). The
expression of $r_{++}$ $=E\{
\alpha 
^{2}\}$ mobilizes the two unknown quantities $\eta $ and $\sigma ^{2},$
which therefore can't be derived from Eq.\ (\ref{spin1/2DensityOperator}),
as one faces a single equation and two unknown quantities.

Our conclusion is different when explicitly using the fact that $\mid \Phi >$
is an RCPS. The probability of getting $+1/2$ as a result of a given trial
is $p_{+}=
\alpha 
^{2}.$ This probability is therefore itself an RV. One may then consider
both its first and second moments, i.e. the second and fourth moments of $%
\alpha $:%
\begin{eqnarray}
E\{p_{+}\} &=&E\{\alpha ^{2}\}=\frac{\int_{0}^{1}\alpha ^{2}e^{-(\alpha
-\eta )^{2}/2\sigma ^{2}}d\alpha }{\int_{0}^{1}e^{-(\alpha -\eta
)^{2}/2\sigma ^{2}}d\alpha }\text{\label{<r^2>}} \\
E\{p_{+}^{2}\} &=&E\{\alpha ^{4}\}=\frac{\int_{0}^{1}\alpha ^{4}e^{-(\alpha
-\eta )^{2}/2\sigma ^{2}}d\alpha }{\int_{0}^{1}e^{-(\alpha -\eta
)^{2}/2\sigma ^{2}}d\alpha }.\text{\label{<r^4>}}
\end{eqnarray}%
The numerical values of $E\{p_{+}\}$ and $E\{p_{+}^{2}\}$ can be estimated
from the experimental (or simulated) data. Thanks to Eq. (\ref{<r^2>}) and (%
\ref{<r^4>}), one now faces a system of two equations, with the two unknown
quantities $\eta $ and $\sigma ^{2},$ which can then in principle be
accessed.

\subsection{Again a spin 1/2, and now two RCPS}

The interest of using the statistical properties of the probabilities (of
the results of measurements) as compared with the use of the operator $r$
associated with an RCPS will now be illustrated by considering a spin $1/2$
described with an RCPS obeying the following equation:%
\begin{equation}
\mid \Phi >=\alpha \mid +>+\sqrt{1-\alpha ^{2}}e^{i\varphi }\mid ->\text{ %
\label{spin1/2RCPS_r-Phi copy(1)}}
\end{equation}%
%
%
%
%
%
%
%
%
%
%
%
%
%
%
%
%
%
%
%
%
%
%
%
%
%
%
%
%
%
%
%
%
%
%
%
%
%
%
%
%
%
%
%
%
%
%
%
%
%
%
%
%
%
%
%
%
%
%
%
%
%
%
%
%
%
%
%
%
%
%
%
%
%
%
%
%
%
%
%
%
%
%
%
%
%
%
%
%
%
%
%
%
%
%
%
%
%
%
%
%
%
%
%
%
%
%
%
%
%
%
%
%
%
%
%
%
%
%
%
%
%
%
%
%
%
%
%
%
%
%
%
%
%
%
%
%
%
%
%
%
%
%
%
%
%
where $\alpha $ and $\varphi $ are real, independent RV, with $0\leq \alpha
\leq 1\,$, $-\pi \leq \varphi \leq \pi \,$, and \ $\varphi $ being uniformly
distributed between $-\pi $ and $+\pi $.\ Then $E\{e^{i\varphi }\}=0\,,$ and 
$r,$ the corresponding density operator with matrix elements $%
r_{lk}=E\{c_{k}^{\ast }c_{l}\}$, is represented by a diagonal matrix in the
standard basis, since:%
\begin{equation}
r_{\text{ }-+}=E\{c_{+}^{\ast }c_{-}\}=E\{\alpha \sqrt{1-\alpha ^{2}}%
e^{i\varphi }\}=E\{\alpha \sqrt{1-\alpha ^{2}}\}E\{e^{i\varphi }\}=0.\text{ %
\label{coherences copy(1)}}
\end{equation}%
%
%
%
%
%
%
%
%
%
%
%
%
%
%
%
%
%
%
%
%
%
%
%
%
%
%
%
%
%
%
%
%
%
%
%
%
%
%
%
%
%
%
%
%
%
%
%
%
%
%
%
%
%
%
%
%
%
%
%
%
%
%
%
%
%
%
%
%
%
%
%
%
%
%
%
%
%
%
%
%
%
%
%
%
%
%
%
%
%
%
%
%
%
%
%
%
%
%
%
%
%
%
%
%
%
%
%
%
%
%
%
%
%
%
%
%
%
%
%
%
%
%
%
%
%
%
%
%
%
%
%
%
%
%
%
%
%
%
%
%
%
%
%
%
%
We then successively consider \ two specific RCPS, $\alpha $ \textit{moreover%
} taking two values only, with the following probabilities:

- in the first RCPS: \{$\alpha =0.45$, $p=0.5$, and $\alpha =0.55$, $p=0.5$%
\},

- in the second RCPS, $\alpha $ again takes two values, but now: \{$\alpha
=0.9$, $p=97/320,$ and $\alpha =0.1$, $p=223/320$\}.

Here, in both cases, $r_{++}=E\{\alpha ^{2}\}=0.2525$: the density operators 
$r$ associated with these two RCPS are identical.

We now consider the first and second moments of the RV $p_{+}$, the
probability of obtaining the value $+1/2$ as a result of the measurement of $%
s_{z}$, in this two level measurement, and this for the first and then for
the second RCPS.

The first moment of $p_{+}$ is $E\{p_{+}\}=E\{\mid c_{+}\mid
^{2}\}=E\{\alpha ^{2}\}$.\ The second moment of $p_{+}$ is $%
E\{p_{+}^{2}\}=E\{\alpha ^{4}\}.$ The numerical value of the first moment of 
$p_{+}$ is therefore already known.\ It is the same for both RCPS: $%
E\{p_{+}\}=0.2525.$ In contrast, the value of the second moment of $p_{+}$
is:

- for the first RCPS, $E\{p_{+}^{2}\}=E\{\alpha ^{4}\}=0.06625625$,

- for the second RCPS, $E\{p_{+}^{2}\}=E\{\alpha ^{4}\}=0.19895$, a value
roughly three times greater than that for the first RCPS.

%
%
%
The two chosen RCPS of\ the spin 1/2 therefore have the same associated
density operator. The probability $p_{+}$ of obtaining the result $1/2$ when
measuring $s_{z}$ is then an RV.\ Whereas the first moment of $p_{+}$ has
the same value for both RCPS, this is not verified with its second moment.%

One may notice that our treatment of quantum systems in states described
with the RCPS concept uses the general postulates of QM, with of course the
exception of the specific one introduced by von Neumann in the treatment of
statistical mixtures (cf. Section \ref{SectionvonNeumannMixture-Postulate}).
When a description with an RCPS is relevant, the use of its statistical
properties is then more powerful than the use of its associated density
operator, a fact which may be translated into the language of the order of
the moments of RV (for its interest in the context of BQPT, see e.g. \cite%
{DevilleY2023QIP}).

\section{About the Landau-Feynman approach\label{SectionLandau-Feynman}}

In the first section of \cite{Landau1927}, entitled "\textit{Coupled systems
in wave mechanics", }Landau wrote: "\textit{A system cannot be uniquely
defined in wave mechanics; we always have a probability ensemble
(statistical treatment). If the system is coupled with another, there is a
double uncertainty in its behaviour". }But an \ operator then introduced
through a \textit{Partial Trace} procedure \textit{in the presence of such a
coupling }does not obey the Liouville-von Neumann equation, and calling it a
density operator may nowadays introduce confusion.\ In Volume III of their
Course (English translation of the second edition, \cite{LandauLifshitzVol3}%
) Landau and Lifshitz first supposed that a\textit{\ "closed system as a
whole is in some state described by a wave function }$\Psi (q,x),$\textit{\
where} $x$ \textit{denotes the set of coordinates of the system considered,
and }$q$ \textit{the remaining coordinates of the system considered". }%
Integrating over the $q$ variables -which corresponds to introducing a
partial trace -, they introduced an operator \textit{which they again called
a density matrix} (thus keeping the difference with its now well-accepted
meaning resulting from the von Neumann approach). Then, in a second step
only, they "\textit{suppose that the system" }(of interest) \textit{"is
closed, or became so at some time". }In Chapter 2 of his \textit{Statistical
Mechanics} \cite{Feynman1972}, Feynman suppressed the possible confusion
resulting from the use of the expression density (or statistical) operator
by both von Neumann and Landau under different assumptions (the possible
existence of a coupling of the system of interest with a second system in
Landau's approach)\textit{.} Feynman considers a system $\Sigma $ composed
of the system of interest, $\Sigma _{1},$ and $\Sigma _{2},$ the rest of the
universe.$\ $He explicitly writes \textit{"it is unknown whether or not the
rest of the universe is in a pure state". }In the following, as in \cite%
{Entropy2017}, $\Sigma _{2}$ will be the\ collection of systems with which $%
\Sigma _{1}$ may interact at the chosen time scale, the whole system $\Sigma 
$ being isolated at this time scale.\ At a time $t_{0}$ when $\Sigma _{1}$
and $\Sigma _{2}$ are uncoupled, $\Sigma _{1}$ and $\Sigma _{2}~$are
separately prepared, each in a pure state. $\Sigma $, the global system, is
therefore in a pure state $\mid \Psi (t_{0})>.$ In a situation when, after
this preparation act, an internal coupling exists between $\Sigma _{1}$\ and 
$\Sigma _{2},$\ and this until some time $t_{1}$,\ one is interested in the
behaviour of $\Sigma _{1}$\ for $t\geq t_{1}$, i.e. once this coupling has
disappeared, at the chosen time scale. Feynman first observes that for $%
t\geq t_{0}$ the whole system obeys the Schr\"{o}dinger equation, and then,%
\textit{\ for}\textbf{\ }$t\geq t_{1},$\textit{\ i.e. after the
disappearance of this internal coupling}, he calculates the mean value of $%
\widehat{O}$ for an arbitrary observable of $\Sigma _{1}$. He first shows
that this mean value at $t_{1}$ is equal to $Tr_{1}\{\rho _{1}(t_{1})%
\widehat{O})\},$ where $\rho _{1}(t_{1})$ $=Tr_{2}\rho (t_{1})$ ($\rho
(t_{1})$ being the projector $\mid \Psi (t_{1})><\Psi (t_{1})\mid $, and $%
\mid \Psi (t_{1})>$ the ket describing $\Sigma $ at $t_{1},$ according to
the Schr\"{o}dinger equation, and $Tr_{1}$ (resp.\ $Tr_{2}$) being \ a trace
calculated over the kets of an orthonormal basis of $\Sigma _{1}$ (resp. $%
\Sigma _{2}$), then that the result keeps true for any time $t>t_{1},$ and
finally that the partial trace $\rho _{1}(t)_{\text{ }}$obeys the
Liouville-von Neumann equation for $t\geq t_{1}$. The use of the Schr\"{o}%
dinger equation for $t\geq t_{1}$ for the establishment of this property
implies that when $t\geq t_{1},$ $\Sigma _{1}$ may be submitted to
time-dependent forces giving birth to a time-dependent Hamiltonian, the
sources of these forces (e.g. an oscillating magnetic field acting on a spin
magnetic moment) being then included in $\Sigma _{1}.$

One may then say that, once $\Sigma _{1}$ is uncoupled from $\Sigma _{2},$
if one is interested in the mean value of the Hermitian operator attached to 
\textit{an observable of }$\Sigma _{1}$\textit{\ only, everything happens as
if }$\Sigma _{1}$\textit{\ were in a statistical mixture}\ described by $%
\rho _{1}(t)$ (it can be verified that $\rho _{1}(t)$ possesses all the
properties of a density operator).

Under the assumptions made, and when $t\geq t_{1}$, the obtained results may
be read by claiming that the system of interest $\Sigma _{1}$ keeps a memory
of its past coupling with $\Sigma _{2},$ and once this has been said the
existence of $\Sigma _{2}$ should be forgotten. The way $\rho _{1}(t_{1})$
is introduced shows that this claimed memory is a manifestation of the
(so-called quantum) correlations created by the $\Sigma _{1}$ - $\Sigma _{2}$
coupling which did exist between $t_{0}$ and $t_{1}$ and created an
entangled state. Of course, someone could take $\Sigma _{2}$ as the system
of interest, and introduce his own so-called reduced density operator $\rho
_{2}(t).$\ But he is not allowed to forget that $\Sigma $ is in a pure state 
$\mid \Psi (t)>,$ and not allowed to suggest that the state of $\Sigma $ is $%
\rho _{1}(t)\otimes \rho _{2}(t),$ generally a statistical mixture.

The situation considered in the Landau-Feynman approach does also mobilize a
density operator $\rho ,$ but is clearly different from the situation
considered by von Neumann. An interest of Feynman's treatment is that, using
already existing quantum postulates, and specifically the fact that the mean
value of $\widehat{O}$ in state $\mid \Psi >$ is $<\Psi \mid \widehat{O}\mid
\Psi >$, it introduces the density operator in a specific situation.
However, by the very link to $\rho $ established in this Feynman-Landau
approach, the postulate made by von\ Neumann is kept, while perhaps masked.
In 1966 B.\ d'Espagnat had called a statistical mixture as defined by von
Neumann a proper mixture, and one imagined from a partial tracing an
improper mixture \cite{d'Espagnat1966}. The 1972 clarification about the
conditions of the use of the Landau approach, by Feynman, did not introduce
the existence of a statistical mixture. After 1972, d'Espagnat however kept
his distinction and the expression \textit{improper mixture} \cite%
{d'Espagnat1999} with, consequently, controversies about the relevance of
this distinction (see e.g. \cite{Kirkpatrick2004}).

QPT often considers a composite system made of the system of interest $%
\Sigma _{1}$ (state space $\mathcal{E}_{1}$) and its environment $\Sigma
_{2} $ (state space $\mathcal{E}_{2}$) and introduces a partial trace over
(a basis of) $\mathcal{E}_{2}$ (see e.g. Ch.\ 8 of \cite{NielsenChuang2005},
and \cite{Branderhorst2009}), with the reservation that introducing a
density operator through a partial Trace is relevant only if the $\Sigma
_{1} $ - $\Sigma _{2}$ coupling has disappeared at the time when this
partial tracing is considered.

Beyond the choice of the words, the important point to be kept is the idea
that the introduction of $\rho $ by von Neumann and the just described
Landau-Feynman partial tracing refer to two distinct physical situations,
the manipulation of an RCPS referring to a third one.

\section{The Zeh problem and the use of higher-order moments\label%
{SectionTheZehProblem}}

In 1927,\ Weyl \cite{Weyl1927}, von Neumann \cite{vonNeumann1927} and Landau 
\cite{Landau1927}\ separately insisted that, in the quantum domain, what
Weyl then called a pure state$~$ (\textit{reiner Fall}) was not the whole
story.\ Von Neumann used the frequentist approach to probabilities then
recently developed by von Mises, and before the publication of Kolmogorov's
work (about Kolmogorov and the frequentist approach, see \cite%
{Kolmogorov1939}). The existence of von Neumann's measurement postulate was
stressed in Section \ref{SectionvonNeumannMixture-Postulate}, but one has to
try and identify the reason of its introduction. In the preface of his 1932
book \cite{vonNeumann1932}, von Neumann wrote that, at the time of its
writing,\textit{\ the relation of quantum mechanics to statistics and to the
classical statistical mechanics }was\textit{\ of special importance}. And 25
years later Fano \cite{Fano1957} noted that, in that time interval, "\textit{%
States with less than maximum information, represented by density matrices }$%
\rho $\textit{, have been considered primarily in statistical mechanics and
their discussion has been influenced by the historical background in this
field}". In the previous development of classical statistical mechanics,
Gibbs had introduced a probability density (within the phase space), used
for the calculations of mean values. In contrast, what corresponds to what
is now called higher-order moments (see e.g. their use in \cite%
{DevilleY2023QIP}) had not been explicitly considered in physics.\
Therefore, when von Neumann introduced his measurement postulate, this he
could implicitly consider not to be responsible for a loss of information as
compared with that contained in the definition of a statistical mixture
through the explicit consideration of the \{$p_{i}$, $\mid \varphi _{i}>$\}
collection.

When examining von Neumann's conception of a statistical mixture from \cite%
{vonNeumann1932}, a first difficulty is the fact that this question occupies
parts of four of its six chapters.\ A second one, for the modern reader,
results from the fact that von Neumann does not use the now standard ket
formalism, introduced seven years later \cite{Dirac1939}. In \cite%
{vonNeumann1932}, von Neumann, having considered the probability content
attached to a pure state, adds (pages 295-296)\ \textit{"the statistical
character may become even more prominent, if we do not even know what state
is actually present \ - - for example when several states }$\phi _{1},$ 
\textit{\ }$\phi _{2},...$\textit{\ with the respective probabilities }$%
w_{1},$\textit{\ }$w_{2},...$($w_{1}\geq 0,$ $w_{2}\geq 0,..$.$%
w_{1}+w_{2}+...=1$) \textit{constitute the description}" of $S$, the quantum
system of interest. He moreover considers\ (page 298)\textit{\ }"\textit{%
great statistical ensembles which consist of many systems }$S_{1}$\textit{\
, . . ., }$S_{N},$ i.\textit{e., N models of S , N large}". Similarly, at
the beginning of his Chapter V, devoted to thermodynamical questions, von
Neumann, extending Gibbs' replica method into the quantum domain, introduces
a mental ensemble of identical systems in which he measures some operator $R$%
, now separating this ensemble into sub-ensembles according to the result of
the measurement. He has started Ch.\ IV of \cite{vonNeumann1932} saying that
in his previous chapter he has "\textit{succeeded in reducing all assertions
of quantum mechanics" }to a formula expressing that the mean value of a
physical quantity $O$ when the system is in the state $\mid \Psi >$ is equal
to a quantity written, with our notations, as $<\Psi \mid \widehat{O}\mid
\Psi >.$ But this he postulated in his Ch.\ III$,$ as the reader may
convince himself: he has first to see the existence of property $E_{2}$ in
page 203 of \ \cite{vonNeumann1932}, and then, in its page 210, to read
that: "\textit{we recognize }$P$\textit{. (or }$E_{2}$\textit{\ .) as the
most far reaching pronouncement on elementary processes}". But von Neumann
has first written: "\textit{We shall now assume this statement P to be
generally valid}" (page 201), and "\textit{We shall now deduce }$E_{1}$%
\textit{. from }$P$\textit{., and }$E_{2}$\textit{. from }$E_{1}$\textit{.}"
(page 203).

Consequently, given a system $\Sigma $ in a statistical mixture described by 
$\rho $ and $\widehat{O}$ attached to an observable $O$ of $\sum ,$ the
assertion that everything should be contained in the expression $E\{\widehat{%
O}\}=Tr\{\rho \widehat{O}\}$ expresses a postulate.

Bell's strong reluctance about the place presently given to measurements in
the foundations of QM \cite{Bell1989} and his question "\textit{Was the
wavefunction of the world waiting to jump for thousands of millions of years
until a single-celled living creature appeared? Or did it have to wait a
little longer, for some better qualified system ... with a PhD?" \ }are
well-known\textit{. }Saying that his question was provocative is a statement
about the question, not an answer. Already in 1970 Zeh \cite{Zeh1970}\
stressed a consequence of that von Neumann postulate (which he called the 
\textit{measurement axiom, }leading to a \textit{circular argument}), when
writing: "\textit{the statistical ensemble consisting of equal probabilities
of neutrons with spin up and spin down in the x direction cannot be
distinguished by measurement from the analogous ensemble having the spins
parallel or antiparallel to the y direction. Both ensembles, however, can be
easily prepared by appropriate versions of the Stern-Gerlach experiment. One
is justified in describing both ensembles by the same density matrix as long
as the axiom of measurement is accepted. However, the density matrix
formalism cannot be a complete description of the ensemble, as the ensemble
cannot be rederived from the density matrix" }\cite{Zeh1970}\textit{.\ }We
call this situation for neutrons proposed by Zeh \textbf{the Zeh problem}.

Zeh introduces a Stern-Gerlach (SG) equipment.\ In their 1922 experiment,
Stern and Gerlach used silver atoms placed in a furnace heated to a high
temperature, leaving the furnace through a hole and propagating in a
straight line. They then crossed an inhomogeneous magnetic field and
condensed on a plate (see \cite{Cohen-Tannoudji2019}, page 394). As they
have no electric charge, they were not submitted to the Laplace force, but
they have an electronic permanent magnetic moment. In a classical approach,
one should then observe a single spot, \textit{whereas two spots} were
observed, which could only be explained, later on, as the result of a
quantum behaviour: a silver atom has a spin $1/2.\ $Zeh considers the random
emission of neutrons by a neutron source.\ It is well-established that a
neutron has a nuclear spin $1/2$ here denoted as $\overrightarrow{s}$\ (it
is usually written as $\overrightarrow{I}$, the symbol $\overrightarrow{s}$
being kept for spins with electronic origin) and a magnetic moment $\mu
=-1.913047$ $\mu _{N}$ ($\mu _{N}:$ nuclear magneton)~proportional to its
spin. The force acting on the magnetic moment of the successive neutrons
deflects them into two well-identified beams, one beam corresponding to the
spin quantum state $\mid z,+1/2>$ and one beam corresponding to the spin
quantum state $\ \mid z,-1/2>.$\ The letter $z$ is reminiscent of the fact
that the field gradient and the force on the spin were directed along $\ z$
in Fig 1, in page 395 of \cite{Cohen-Tannoudji2019}. As the neutrons are
emitted one by one (no interaction between them),\ interact only with the
magnetic field before being collected on the plate, and are not each one
identified when leaving the furnace, but are only counted when arriving on
the plate, with the same total number $N/2$ in the two packets, one may say
(strictly speaking, in the limit $N$ $\longrightarrow \infty $) that one
prepared the following (von Neumann) statistical mixture: $\mid +z,1/2>,$ $%
\frac{1}{2},$ $\mid -z,1/2>,$ $\frac{1}{2}.\ $This mixture is the one
compatible with the SG equipment in reference \cite{Cohen-Tannoudji2019}.\
Following up the question from Zeh in \cite{Zeh1970}, we now consider a spin 
$1/2$,\ and successively its state in:

\begin{center}
Mixture 1: $\mid +x>,$ $1/2$ and $\mid -x>,$ $1/2,$

Mixture 2: $\mid +y>,$ $1/2$ and $\mid -y>,$ $1/2,$
\end{center}

$\hspace{-0.5cm}\mid +x>$ and $\mid -x>$ being the eigenkets of $s_{x}$ for
the values $+1/2$ and $-1/2$ respectively, and $\mid +y>$ and $\ \mid -y>$
the eigenkets of $s_{y}$ for the values $+1/2$ and $-1/2$ respectively.

The density operator associated with both mixtures is $\rho =I/2$ ($I$:
identity operator in the state space of the spin). We decide to forget the
existence of the von Neumann measurement postulate, which suggests that both
mixtures\ are the same, and therefore discourages us from doing what
follows. We choose to use, instead of the $\rho $ formalism, the very
definition of these mixtures. And, in order to try and clarify the Zeh
problem, our previous use of moments in the presence of an RCPS here
suggests us to use moments of an arbitrary order (and not only the mean
value) of a well-chosen RV. Just before the plate, at the level of each
arriving beam, we introduce an equipment able to measure the $\mathbf{s}_{%
\mathbf{x}}$ component of each neutron, and to store the result. Von Neumann
told us that the mean value of the result of this measurement, written in
the Dirac formalism, is:%
\begin{eqnarray*}
\frac{1}{2} &<&+x\mid s_{x}\mid +x>+\frac{1}{2}<-x\mid s_{x}\mid -x>\text{
for mixture 1} \\
\frac{1}{2} &<&+y\mid s_{x}\mid +y>+\frac{1}{2}<-y\mid s_{x}\mid -y>\text{
for mixture 2}
\end{eqnarray*}%
One can interpret these results as the mean value (over all pure states that
compose the considered mixed state) of a random variable which we denote as $%
X,$ and which is defined as being itself the \textit{mean }value taken by $%
s_{x}$ when the spin is in a given pure state. Its name $X$ recalls us that
it mobilizes the $x$ component of the spin. In the specific case of a pure
state $\mid \varphi >,$ $X$ takes the value $<\varphi \mid s_{x}\mid \varphi
>.$

For any value of the non-negative integer $n,$ $\mu _{n},$ the $nth$ moment
\ of $X$ has the value for mixture 1:%
\begin{equation*}
\text{mixture 1}\text{:}\text{\ }\mu _{n}(X)=\frac{1}{2}(<+x\mid s_{x}\mid
+x>)^{n}+\frac{1}{2}(<-x\mid s_{x}\mid -x>)^{n}
\end{equation*}

\begin{equation*}
\hspace{-3cm}\hspace{1.3cm}=\frac{1}{2}(\frac{1}{2})^{n}+\frac{1}{2}(-\frac{1%
}{2})^{n}
\end{equation*}%
\textit{Therefore, in statistical mixture 1, any odd moment of }$X$ \textit{%
has a value equal to }$0,$\textit{\ and any even moment (}$n$\textit{\ even)
is equal to }$1/2^{n}.$

Considering now mixture $2$, the $nth$ moment \ of $X$ has the value:%
\begin{equation*}
\text{mixture 2: }\mu _{n}(X)=\frac{1}{2}(<+y\mid s_{x}\mid +y>)^{n}+\frac{1%
}{2}(<-y\mid s_{x}\mid -y>)^{n}
\end{equation*}

We recall the developments of $\mid +y>$ and $\mid -y>$ within the standard
basis:%
\begin{equation*}
\mid +y>=\frac{\mid +>+i\mid ->}{\sqrt{2}}\text{ \ \ and \ \ \ }\mid -y>=%
\frac{\mid +>-i\mid ->}{\sqrt{2}}\text{ }
\end{equation*}%
The quantity $<+y\mid s_{x}\mid +y>$ is equal to zero, as the diagonal
quantities $<+\mid s_{x}\mid +>$ and $<-\mid s_{x}\mid ->$ are both equal to 
$0$, and the sum of the interference terms is equal to zero.\ The same
result is obtained for $<-y\mid s_{x}\mid -y>.$

\textit{Therefore, in statistical mixture 2, any moment of }$X$\textit{\ is
equal to 0.}

One guesses that if, in contrast, the same mixtures being considered, one
measures $s_{z}$ instead of $s_{x},$ and one then introduces the RV $Z$,
defined in the same way as $X$ (and which, of course, has nothing to do with 
$Z,$ the direction of a\ magnetic field), the difference found with the
moments of $X$ should disappear with $Z,$ since the choice of $s_{z}$
introduces a new symmetry, and an inability for the new RV $Z$ to
distinguish between the two mixtures through the use of the moments of $%
s_{z} $. We choose to examine this question explicitly. $Z,$ the new RV, is
defined through the way already used for $X,$ $s_{x}$ measurements being
replaced by $s_{z}$ measurements.\ One first considers the values of the
moments of $Z$ when the spin is in mixture $1.$\ The developments of $\mid
+x>$ and $\mid -x>$ in the standard basis are respectively:%
\begin{equation*}
\mid +x>=\frac{\mid +>+\mid ->}{\sqrt{2}}\text{ \ \ and \ \ \ }\mid -x>=%
\frac{\mid +>-\mid ->}{\sqrt{2}}\text{ }
\end{equation*}%
The value of $Z$ in the pure state $\mid +x>$, i.e. $<+x\mid s_{z}\mid +x>,$
when $\mid +x>$ is developed in the standard basis, is obtained as the sum
of its interference terms, each equal to zero, and of the diagonal terms,
the sum of their contributions being equal to $0.\ $Therefore $<+x\mid
s_{z}\mid +x>=0.$\ For the same reason, $<-x\mid s_{z}\mid -x>=0.$
Therefore, any moment of $s_{z}$ in mixture 1 has a value equal to $0.\ $%
Following the same approach, one gets the same result for $Z$ and mixture 2.$%
\ $As expected, considering measurements of $s_{z}$ and the moments of $Z,$
one is unable to establish any difference between Zeh mixtures $1$ and $2.$
This result however does not change the previous conclusion, which
corresponds to a sufficient condition: using two well-chosen mixtures -those
introduced by Zeh- possessing the same density operator, we have been able
to introduce a well-chosen RV, which we called $X,$ related to results of
measurements of a well-chosen spin component, namely $s_{x},$ and such that
at least one of the moments of $X$ had a different value in the two Zeh
mixtures.

In summary, in his 1970 paper focused on the spin of neutrons and
Stern-Gerlach equipments, and on two statistical mixtures chosen so that
both mixtures have the same denstiy operator $\rho =I/2$\ , Zeh observed
that the description with $\rho $ should not tell the whole story for these
mixtures, since it forgets the initial preparation process of these
mixtures.\ We have just: 1) decided to ignore the von Neumann measurement
postulate (cf.\ Section \ref{SectionvonNeumannMixture-Postulate}), 2)
introduced a well-chosen spin-operator, $s_{x},$ and an RV denoted as $X$
and compatible with what one usually says in QM about the mean value of an
observable in the presence of a statistical mixture, 3) established that the
even moments of X have different values in mixture 1 and in mixture 2.\ This
result allows us to say that, contrary to what is claimed when assuming the
von Neumann measurement postulate, these two mixtures should be
distinguished.\ This result is a sufficient property: when two mixtures have
the same density matrix, once the von Neumann postulate has been given up,
one should consider the very definition of a given statistical mixture, and
use e.g. a well-defined RV linked to this mixture, and its moments.\ The
associated density operator, certainly an important tool, does not
necessarily contain the whole information contained in the mixture \{$p_{i}$%
, $\mid \varphi _{i}>$\}, which confirms an intuition from Zeh.

\section{Conclusion\label{SectionConclusion}}

Quantum Information Processing (QIP) is expanding its own place within
Quantum Electronics. In the development of Blind Quantum Source Separation
(BQSS) and Blind Quantum Process Tomography (BQPT), the use of the concept
of a random-coefficient pure state (RCPS) has been found useful, within
standard Quantum Mechanics (QM) and its Hilbert space framework. An RCPS\
has to be clearly distinguished from a statistical mixture - a quantum
isolated physical system in different pure states, each one with a given
probability - which, as a consequence of von\ Neumann's work, uses the
density operator $\rho $ as the formal tool for the treatment of its
statistical properties. In the present paper, an experiment leading to the
existence of an RCPS was presented. When a system is described with an RCPS,
probabilities of results of measurements of observables become themselves
Random Variables (RV), presenting an informative content through their
statistical laws. With a given RCPS one may associate a single,
well-defined, density operator, and with a given density operator one may
associate more than one random-coefficient pure state.\ An instance of a
spin 1/2 in an RCPS with two unknown parameters was introduced, and it was
shown that information obtained from $s_{z}$ measurements allows the
determination of these parameters using known properties of the statistical
laws obeyed by these probabilities, while that determination is impossible
using the density operator associated with this RCPS. It was stressed that
the use of the $\rho $ formalism in the description of a statistical mixture
rests upon a postulate. It was also shown that the use of what we called the
Landau-Feynman approach, in a situation well identified by Feynman, makes an
implicit use of this postulate, by its very use of the $\rho $ operator. In
the presence of a von Neumann mixture, stimulated by the existence a 1970
paper from Zeh, by our use of the moments of RV in the presence of an RCPS,
and by the reason which historically led von\ Neumann to introduce his
density operator, we gave up the von Neumann measurement postulate and then
established a sufficient condition allowing us to say that, in the presence
of a von Neumann statistical mixture, the\ exclusive use of the density
operator may imply a loss of some information contained in the very
definition of that mixture, which can be kept if one considers the moments
of a well-chosen RV associated with results of measurements of observables.

The present paper therefore contains two main contributions with respect to
quantum systems and algorithms, that are of importance for the Quantum
Electronics Section of this Journal. The first one relates to the
representation of the quantum states manipulated in these systems and
algorithms, and to associated measurements: whereas the usual representation
of mixed states is restricted to the use of the density operator formalism,
we showed that their original representation as a set of
(deterministic-coefficient) pure states and associated probabilities,
together with adequate measurements, may allow one to extract more
information about them than the one contained in their density operator,
which in turn may yield more powerful information processing capabilities.
Moreover, our second contribution relates to the exploitation of results of
measurements performed for the mixed states considered above, or for the
random-coefficient pure states that we introduced in our previous papers and
that we further analyzed here: we showed that higher-order statistics of
random variables associated with both types of states allow one to extract
more information about these states and hence to extend quantum information
processing capabilities.

\section{Appendix\label{SectionAppendix}}

A justification of some properties obeyed by the operator $r$\ associated
with a random-coefficient pure state, introduced in Subsection \ref%
{SubsectionFromrRCPStoRho}, is given here, using the notations of that
section. $r$ acts on the (deterministic-coefficient) states of $\mathcal{E}$%
, the state space of $\Sigma .$\ For instance, when considering $r\widehat{O}%
\mathbf{\mid }\Psi _{s}\mathbf{>,}$ $r$ acts on the
(deterministic-coefficient) state $\widehat{O}\mathbf{\mid }\Psi _{s}\mathbf{%
>}$ resulting from the action of $\widehat{O}$ on the
(deterministic-coefficient) state $\mathbf{\mid }\Psi _{s}\mathbf{>.}$

Property 1: hermiticity of $r$. If $X$ is a complex random variable with $%
X=A+iB,$ $A$ and $B$ being real random variables, then $(E\{X\})^{\ast
}=E\{A\}-iE\{B\}=$ $E\{X^{\ast }\}\ $(expec\-tation and complex conjugation
commute).\ Therefore, considering the matrix with elements%
\begin{equation}
r_{lk}=E\{c_{k}^{\ast }c_{l}\}
\end{equation}%
in the chosen basis, then $r_{kl}^{\ast }=(E\{c_{l}^{\ast }c_{k}\})^{\ast
}=E\{c_{l}c_{k}^{\ast }\}=r_{lk}.\ $Therefore, the matrix with elements $%
r_{lk}$ and the operator $r$ are Hermitian.

Property 3: $<u\mid r\mid u>=<u\mid \sum_{i}p_{i}\mid i><i\mid
u>=\sum_{i}p_{i}\mid <i\mid u>\mid ^{2}\geq 0.$

Property 4: $\Sigma $ has been assumed to be isolated, with Hamiltonian $%
\mathcal{H}$.\ Once a (deterministic-coefficient) pure state $\mid \Psi
_{s}> $ has been defined, its time evolution is well-defined, following the
Schr\"{o}dinger equation. If an RCPS is defined at some time $t_{r}$ ($r:$
reference), as $\mid \Psi >=\sum_{k}c_{k}\mid k>,$ its time behaviour is
therefore defined by this Hamiltonian and by the probability laws associated
with the random variables $c_{k}$ defined at time $t_{r}.$ Consequently:%
\begin{eqnarray}
\frac{d}{dt}r &=&\sum_{k,l}\frac{dr_{kl}}{dt}\mid k><l\mid =\sum_{k,l}\frac{%
dE\{c_{l}^{\ast }c_{k}\}}{dt}\mid k><l\mid \\
&=&\sum_{k,l}E\{\frac{d(c_{l}^{\ast }c_{k})}{dt}\}\mid k><l\mid \\
&=&\sum_{k,l}E\{\frac{dc_{l}^{\ast }}{dt}c_{k}+c_{l}^{\ast }\frac{dc_{k}}{dt}%
\}\mid k><l\mid .
\end{eqnarray}

Since $i\hslash \sum_{k}(dc_{k}/dt)\mid k>=\sum_{k}c_{k}\mathcal{H}\mid k>,$
then%
\begin{eqnarray}
i\hslash \frac{d}{dt}r &=&\sum_{k,l}E\{-c_{k}c_{l}^{\ast }\mid k><l\mid 
\mathcal{H}+c_{l}^{\ast }c_{k}\mathcal{H}\mid k><l\mid \} \\
&=&-r\mathcal{H}+\mathcal{H}r=[\mathcal{H},r].
\end{eqnarray}

$r$ therefore obeys the Liouville-von Neumann equation.

%
%

%
%
%
%
%
%


\section*{Statements and declarations}

The authors declare to have no financial or non-financial conflict of
interest. 

\end{document}